\documentclass[conference]{IEEEtran}
\IEEEoverridecommandlockouts
\usepackage{cite}
\usepackage{amsmath,amssymb,amsfonts}
\usepackage{algorithmic}
\usepackage{graphicx}
\usepackage{textcomp}
\usepackage{xcolor}
\def\BibTeX{{\rm B\kern-.05em{\sc i\kern-.025em b}\kern-.08em
    T\kern-.1667em\lower.7ex\hbox{E}\kern-.125emX}}

\usepackage{microtype}
\usepackage{booktabs}
\usepackage{graphicx}
\usepackage{amssymb}
\usepackage{multirow}
\usepackage{enumitem}

\usepackage{fancyhdr}
\usepackage{tikz}
\newcommand\copyrighttext{%
    © 2024 IEEE.  Personal use of this material is permitted.  Permission from IEEE must be obtained for all other uses, in any current or future media, including reprinting/republishing this material for advertising or promotional purposes, creating new collective works, for resale or redistribution to servers or lists, or reuse of any copyrighted component of this work in other works.}
\newcommand\copyrightnotice{%
\begin{tikzpicture}[remember picture,overlay]
\node[anchor=south,yshift=65pt] at (current page.south) {\fbox{\parbox{\dimexpr\textwidth-\fboxsep-\fboxrule\relax}{\copyrighttext}}};
\end{tikzpicture}%
}

\begin{document}
\copyrightnotice

\title{Language-based Audio Retrieval with \\ Co-Attention Networks
}

\author{\IEEEauthorblockN{Haoran Sun, Zimu Wang, Qiuyi Chen, Jianjun Chen, Jia Wang, Haiyang Zhang$^*$\thanks{$^*$Haiyang Zhang is the corresponding author.}}
\IEEEauthorblockA{
\textit{School of Advanced Technology, Xi'an Jiaotong-Liverpool University, Suzhou, China}\\
\{haoran.sun20, zimu.wang19, qiuyi.chen2002\}@student.xjtlu.edu.cn\\
\{jianjun.chen, jia.wang02, haiyang.zhang\}@xjtlu.edu.cn}
}

\maketitle

\begin{abstract}
In recent years, user-generated audio content has proliferated across various media platforms, creating a growing need for efficient retrieval methods that allow users to search for audio clips using natural language queries. This task, known as language-based audio retrieval, presents significant challenges due to the complexity of learning semantic representations from heterogeneous data across both text and audio modalities.
In this work, we introduce a novel framework for the language-based audio retrieval task that leverages co-attention mechanismto jointly learn meaningful representations from both modalities.
To enhance the model’s ability to capture fine-grained cross-modal interactions, we propose a cascaded co-attention architecture, where co-attention modules are stacked or iterated to progressively refine the semantic alignment between text and audio. 
Experiments conducted on two public datasets show that the proposed method can achieve better performance than the state-of-the-art method. Specifically, our best performed co-attention model achieves a 16.6\% improvement in mean Average Precision on Clotho dataset, and a 15.1\% improvement on AudioCaps.
\end{abstract}


\section{Introduction}
In recent decades, the generation and dissemination of audio content on multimedia platforms have significantly increased, driving the need for effective large-scale audio retrieval methods.
Natural language texts can be an efficient medium to summarize details of video and audio recordings with merely a few words, representing information about data beyond any fixed taxonomies. As a result, a strong relationship is anticipated between textual captions and audio signals, as both encapsulate important environmental attributes and source information \cite{xie2022dcase,cscwd2024_fuyu}. 

Language-based multimodal retrieval has received extensive research in recent years, most of which pays their attention to the visual domains \cite{qu2021imagetext,cheng2020coatt}.
% , such as the Co-Attention Network proposed for audio-visual representation \cite{cheng2020coatt}. 
In contrast, the text-audio retrieval task has fewer attentions in the existing literature. 
% Most existing work on text-audio retrieval focus on investigating different methods to encode text and audio resources \cite{xie2022dcase} 
Generally, the major challenge for such task is to learn embeddings from different modalities into a common vector space with invariant semantic inputs, which is much more challenging for long-sequenced audio recordings \cite{Lou2022AudioTextRI}. Current solutions to address such issue are to use attention mechanism  to learn fine-grained representations of both text and audio \cite{cscwd2024_fuyu,xin2023taratt}. It can enable the model to assign attention weights to different parts of the sequences and update them dynamically, so that the model can concentrate on different part of the text and audio semantically. However, most existing studies focus on performing attention operations within individual modalities rather than investigating the interconnections between features from different modalities \cite{xie2022dcase,lai2022resnet}. Such approaches may negatively impact the discovery of potential cross-modal information and the overall performance of retrieval tasks \cite{Nguyen2018ImprovedFO}. Moreover, in research focused on designing cross-modal attention pooling methods to enhance task performance \cite{cheng2021videotext,lou2022atr}, they often neglect the in-depth cascade modeling \cite{xin2023taratt,Koepke2021AudioRW}.

Inspired by the use of co-attention mechanism in visual and textual modalities \cite{cheng2020coatt,Nguyen2018ImprovedFO}, this paper introduce a novel framework for language-based audio retrieval with co-attention network to facilitate information exchange between text and audio. 
The co-attention mechanism enables simultaneous attention to multiple inputs, capturing correlations and relationships across different modalities. It has been adapted to the relationship modeling between two sequences, including questions answering \cite{Santos2016AttentivePN} and recommendation systems \cite{Tay2018recommendation}, and is supposed to be practical in the language-based audio retrieval task.
Different from previous work that only attend text to corresponding audio segments in one direction \cite{cscwd2024_fuyu,xin2023taratt}, our approach can jointly attend individual words and audio segments to capture more semantic representation for both modalities. Moreover, to overcome the limitation of coarse multimodal interaction that shallow models suffers, we propose to investigate two types of deep co-attention networks, called stacking and iterating, for the language-based audio retrieval task, so that the co-attention module can be cascaded in depth to gradually refine the representation of both text and audio.

The contribution of this paper is summarized as follows:

\begin{itemize}
    \item We propose a novel framework for language-based audio retrieval tasks using co-attention mechanism, which can capture the semantic correlation between individual words and audio segments. 
    \item We propose to cascade the co-attention model in depth to capture richer multimodal information, so as to further improve the retrieval performance. Two types of deep co-attention models: stacking model and iterating model, have been investigated. 

    \item Experiments conducted on two public datasets demonstrate that our method can achieve better performance. Comparing with the state-of-the-art models, our best performed model (iterating model) can achieve a 16.6\% improvement in mean Average Precision (mAP) on Clotho dataset and a 15.1\% improvement on AudioCaps.
\end{itemize}

\section{Related Work}

\subsection{Text-Audio Retrieval}

The text-audio retrieval tasks mainly focus on retrieving target audio clips given a natural language query. Early works in this domain managed to employ Euclidean distance to measure the distance between queries and audio samples after compressing and transforming both of them into the same latent vector space \cite{Heln2007QueryBE}. However, one limitations for such method is that it heavily relies on the length of input vectors and thus influencing the precision of the retrieval results. To overcome this issue, Slaney et al. \cite{slaney2002sar} propose to apply neural networks to project the representation of both modalities in the hierarchical common space, and use Kullback–Leibler (KL) distance to calculate the similarity.
Levy and Sandler \cite{levy2009music} introduce a joint codebook for social tags and audio tags, and utilize probabilistic Latent Semantic Analysis techniques to address these tasks. Recent studies also explore the audio-text alignment approach for text-audio retrieval, inspired by the work of \cite{karpathy2015dvsa} and further developed by \cite{Elizalde2019crossmodal}. This approach involves learning features using networks after aligning both audio and textual data in the same latent space by incorporating two embedding sequences into the input data. They also propose further application of result ranking based on the joint embeddings.
Recent approaches to text-audio retrieval tasks include self-supervised learning and zero-shot predictions. A promising middle ground is learning semantic knowledge through natural language supervision, as proposed by Elizalde \textit{et. al} \cite{clap}, which demonstrated superior performance in their work.

\subsection{Co-Attention Mechanism}
Transformers \cite{attention} are rapidly emerging as a promising technique for multi-modal tasks. 
Co-attention networks have become a widely adopted mechanism in Transformer-based models \cite{vit}. This approach traces its origins to the work of \cite{Xiong2016DynamicCN} in the field of Question Answering (QA). Several studies have explored the application of co-attention networks in the audio and textual domains.
For example, Liu \textit{et al.} \cite{Liu2021CoattentionNW} introduced a co-attention network for text classification tasks with joint label embeddings to develop related features, where the co-attention mechanism facilitates the combination and construction of relationships between text and label information, allowing the computation of sequence data and the generation of augmented representations for both data inputs. Their results demonstrate a notable performance improvement over previously widely-used models. 
Similar structures are also seen in recommendation systems \cite{Tay2018recommendation}, where co-attention mechanisms \cite{Zhang2017AttentiveIN} are utilized to learn representations without fine-grained interaction modeling. However, it is important to note that the performance of this technique may be affected by noise and dataset size \cite{huang2023makeanaudio}. The applicability of this approach in the audio domain is further verified by \cite{cheng2020coatt}, who used the co-attention module to address the problem of audio-visual synchronization and successfully learned cross-modal semantic information in their downstream tasks.

\begin{figure*}[t]
    \centering
    \includegraphics[width=\textwidth]{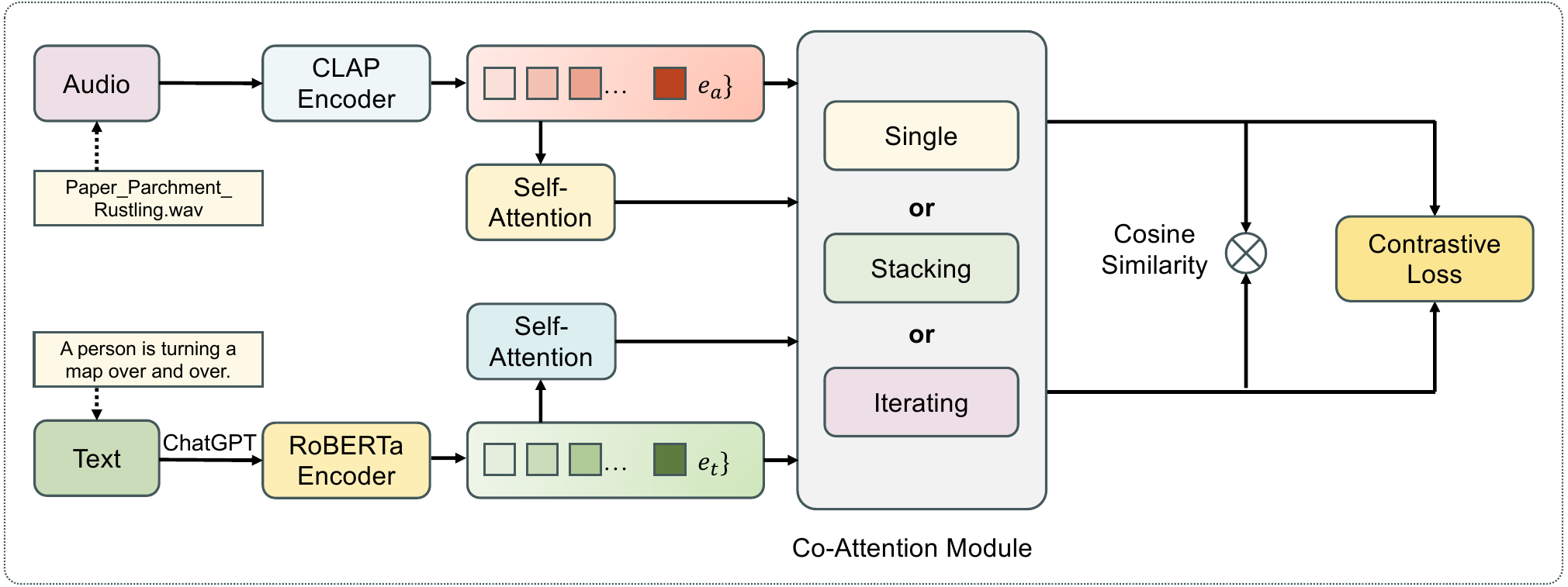}
    \caption{Brief illustration of our proposed model, which includes a GPT generate component for text argumentation, both audio and text self-attended components, a co-attention module and a model training and fusion component.}
    \label{fig:model}
\end{figure*}

\begin{table}[]
    \centering
    \caption{Example of GPT generated captions.}
    \label{tab:gpt}
\begin{tabular}{l|l}
\toprule
\multicolumn{1}{c|}{Original Captions}                          & \multicolumn{1}{c}{Augmented Captions by GPT}  \\ \midrule
 & Paper sheets pass through a printer.           \\
     A printer printing    & Mechanical device prints numerous pages.       \\
     off multiple pages    & Printing machine produces multiple papers.     \\
     of paper    & Sheets of paper emerge from a printing device. \\
                                                                & Printer generates consecutive printed pages.  
\\ \bottomrule
\end{tabular}
\end{table}

\section{Methodology}

\subsection{Problem Definition}

We define the language-based audio retrieval task as follows: for each query text $t$, our goal is to retrieve a ranked list of \(k\) audio recordings (each audio is denoted as $a$) that are most relevant to the query.
The audio recordings and textual captions are encoded by pre-trained language models. In this paper, we use CLAP \cite{clap} to encode  audios and RoBERTa \cite{roberta} to encode texts respectively, 
% and projected into a sharing vector space, 
where the text representation is denoted as \(e_{t}\in \mathbb{R}^{D}\)  and the audio representation is denoted as \(e_{a}\in \mathbb{R}^{D\times N}\) ($D$ represents the dimension of the feature and $N$ denotes the number of audio frames).

In this work, we propose a novel framework using co-attention network for the language-based retrieval task, as illustrated in Figure 1. The architecture of this framework consist of three parts: audio encoder, text encoder and the co-attention module that facilitate the interaction between the two modalities. More specifically, we investigate three different ways utilizing the co-attention network: 1) single module which is a one layer co-attention network; 2) stacking module which cascade the single module in depth with stacking manner; 3)  iterating module which cascade the single module in depth with iterating manner.

\subsection{Audio and Text Encoders}
The encoding process is responsible for transforming raw audio and text inputs into the embedding space that can be further processed by subsequent co-attention and contrastive learning modules.
In order to enhance the language diversity and reduce the bias caused during the annotation process, we use ChatGPT to augment the caption \cite{cscwd2024_fuyu}.
Table \ref{tab
} presents an example where ChatGPT is used to generate five alternative captions based on the original one. Cosine similarity is then applied to select the most relevant caption, forming new audio-text pairs.

For audios, the CLAP encoder is used to capture the rich acoustic features of the audio input while the captions are encoded by RoBERTa. Given the input audio $a$ and text $t$, 
we can obtain their embedding \(e_a\in\mathbb{R}^{N\times d}\), \(e_{t}\in\mathbb{R}^{1\times d}\) with the function below:
\begin{equation}
    e_a = F_a(a), \quad e_t = F_t(t),
\end{equation}
where $d$ represents the dimension of the feature, $N$ denotes the number of audio frame and \(F(\cdot)\) is corresponding encoders. Hence, we are able to extract the key features from the audio and text data and make further training process more effective.

\subsection{Single Co-attention Module}
\begin{figure}[t]
    \centering
    \includegraphics[width=0.27\textwidth,keepaspectratio]{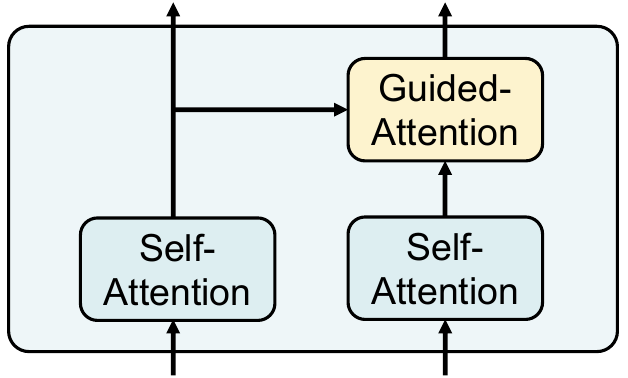}
    \caption{Brief illustration of the single co-attention module, which includes two self-attention modules and a guided-attention components.}
    \label{fig:single}
\end{figure}

The attention mechanism is a crucial technique that allows models to selectively focus on key elements within an input sequence while dynamically adjusting weights. Its widespread adoption in multimodal tasks stems from its ability to effectively manage features across various modalities.
The co-attention mechanism enhances this process by facilitating effective information fusion across different modalities. In this paper, we propose utilizing the co-attention mechanism to capture richer semantic information from both audio and text sources.
As shown in Figure \ref{fig:single}, a co-attention module (also referred to as a single module) consists two self-attention (SA) modules for both modalities to concentrate on related representations and making projections for them, and a guided-attention (GA) module that receives the output of the SA modules. The projection for an audio embedding \(e_a\in\mathbb{R}^{N\times d}\) and a text embedding \(e_{t}\in\mathbb{R}^{1\times d}\) on the dimension of size \(d\) can be defined as:

\begin{equation}
    \begin{gathered}
    Q_{a}=W_{Q_{a}}\cdot LN(e_{a}), \quad Q_{t}=W_{Q_{t}}\cdot LN(e_{t}),
    \\ K_{a}=W_{K_{a}}\cdot LN(e_{a}), \quad K_{t}=W_{K_{t}}\cdot LN(e_{t}),
    \\ V_{a}=W_{V_{a}}\cdot LN(e_{a}), \quad V_{t}=W_{V_{t}}\cdot LN(e_{t}),
    \end{gathered}
\end{equation}
where \(LN\) stands for a normalization layer, \(W_{Q_a}\), \(W_{K_a}\), \(W_{V_a} \) and \(W_{Q_t} \) , \(W_{K_t} \) , \(W_{V_t} \)  are learnable projection weight matrices for audio and textual sections, respectively. The projection transforms the embeddings into queries \(Q \in\mathbb{R}^{1\times d}\), key-value pairs \(K\in\mathbb{R}^{N\times d}\) and value matrices \(V\in\mathbb{R}^{N\times d}\) for both sides. 
To obtain attention weights, we compute the dot products of the query with all keys, then normalize each of them by \(\sqrt{d_k}\) before employing the softmax function on it. Afterwards, we calculate the aggregate feature \(f_a\) and \(f_t\) according to the learned weights and the query and key-value pairs:
\begin{equation}
    \begin{aligned}
        f=Attention(Q,K,V)=softmax\biggl(\frac{Q\cdot K^{T}}{\sqrt{d_{k}}}\biggr)V.
    \end{aligned}
\end{equation}

\begin{figure}[t]
    \centering
    \includegraphics[width=\linewidth]{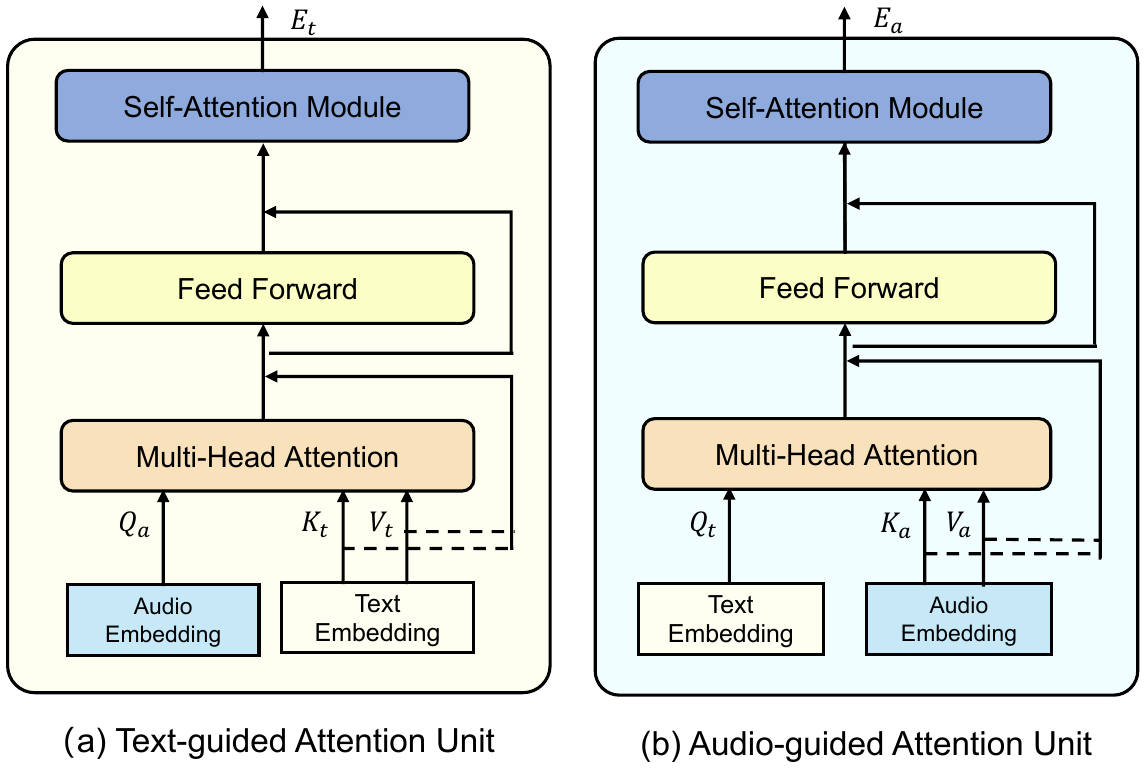}
    \caption{Brief illustration of the guided-attention components for both modalities.}
    \label{fig:guided-attention}
\end{figure}

To achieve better interactions and feature combinations, we modify the original self-attention module for the guided-attention module, as shown in Figure \ref{fig:guided-attention}. The text-guided attention unit receives a query from the audio segment embeddings combining with the original key-value pairs and value matrices from the textual section while, on the contrary, the audio-guided attention unit receives textual queries to aggregate the original audio frame information.
Taking the text modality unit as an example, it seeks to integrate the features of the audio clip into the text.  In this case, the text queries \(Q_t\) are replaced with the audio encoder queries  \(Q_a\in\mathbb{R}^{1\times d}\). Consequently, the modified dot-product similarity \(gf_{t|a} \in \mathbb{R}^{1\times d}\) for this block can be represented as follows:
\begin{equation}
    gf_{t|a}=Att\left(Q_a,K_t,V_t\right)=softmax\left(\frac{Q_a\cdot K_t^T}{\sqrt{d_{k_t}}}\right)V_t,
\end{equation}

The value-projected embedding contains the context of captions, which is attended relying on the queries of audio clipping frames. Similarly, to obtain a more comprehensive representation of the textual embedding, we utilize the multi-head attention unit in this process. The corresponding final output feature \(GF_{t|a} \in \mathbb{R}^{h\times d}\) of the text-guided co-attention block after the \(h\)-multi-head attention operation is:
\begin{align}
GF_{t|a} &= MHA(Q_{a},K_{t},V_{t}) \nonumber \\
    &= Concat(head_{1},head_{2},...,head_{m})W_{o}, \\
head_{i} &= Attention(Q_{a}W_{i}^{Q},K_{t}W_{i}^{K},V_{t}W_{i}^{V}),
\end{align}
where \(W_{o} \in \mathbb{R}^{m\times d\times d_{m}}\) is the set of the weights and the projection matrices of the \(i\)-th head are \(W_{i}^{Q},W_{i}^{K},W_{i}^{V} \in \mathbb{R}^{d\times d_{m}}\). The final output of this can be an aggregated text embedding after transforming in the layer normalization layer and relies on the given audio conditions. Moreover, the output feature of audio-guided co-attention components can be represented in a similar way.

Overall, the co-attention module combining the self-attention and guided-attention units is able to lay more emphasis on most relevant representations while ignoring the unimportant information as much as possible, thus strengthening the interaction and relation between audio and text data and improving its effectiveness.

\subsection{Stacking and Iterating Modules}

\begin{figure}[t]
      \centering
      \includegraphics[width=0.9\linewidth]{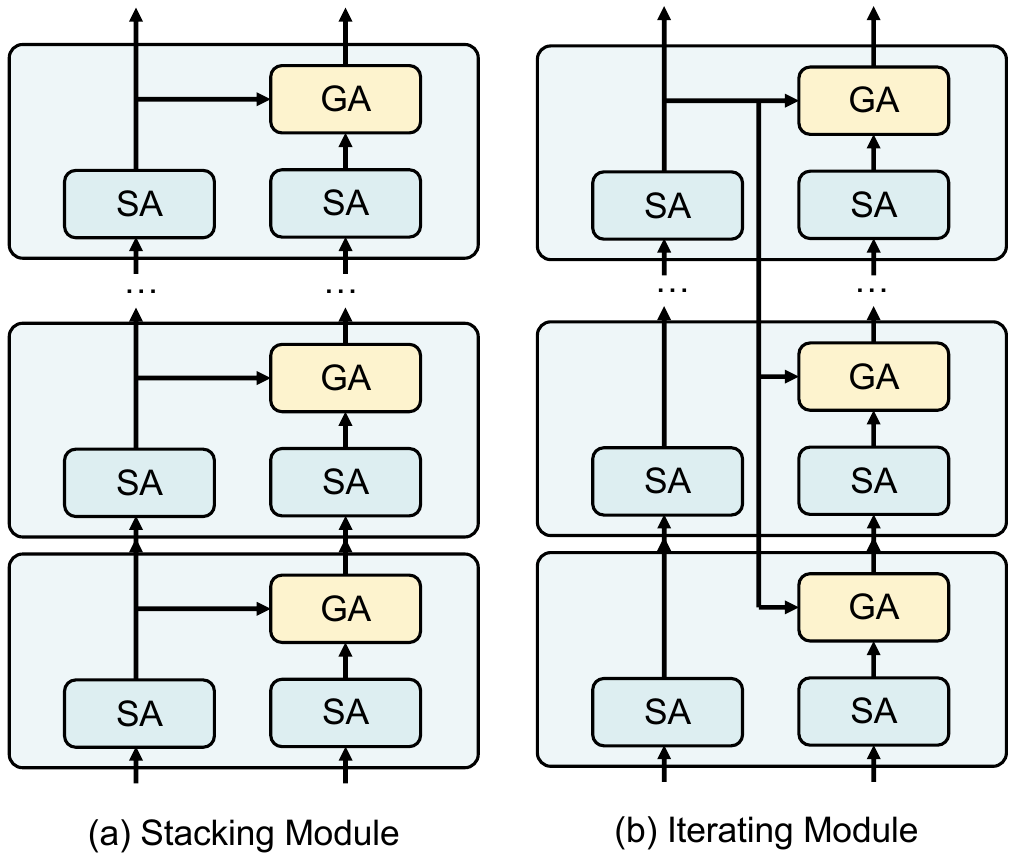}
      \caption{The structure of stacking and iterating modules. stacking module simply stacks the attention components and iterating module trains the network hierarchically by iterating the embedding only requiring self-attention first.}
      \label{fig:stack-iterate}
    \end{figure}

Based on the single co-attention module, we further design the stacking and iterating modules to conduct deep co-attention learning for the audio-text retrieval tasks with multiple layers. Figure \ref{fig:stack-iterate} shows the brief structure of the stacking and iterating modules with $n$ layers in depth. In this section, we use the process of audio-guided co-attention learning as an example to introduce the two units.

As shown in Fugure \ref{fig:stack-iterate}(a), the stacking module simply stacks the co-attention units to $n$ layers and iterates the features in each layer. Let \(A\) and  \(T\) be the original audio and text feature input and \(A_n\), \(T_n\) be the final output of the deep co-attention module, which also represents the output of the \(n\)-th layer in the network. Then, the iteration of the \(k\)-th layer in the module can be defined as:
\begin{equation}
    [A_{k},T_{k}]=Stacking_{k}([A_{k-1},T_{k-1}]).
\end{equation}

Figure \ref{fig
}(b) shows that the model first focuses on obtaining the output features from the textual representation, denoted as \(T_n\). These features are subsequently transferred to the audio section by supplying relevant queries during each iteration of the audio-guided attention module layers. The audio embeddings output \(A_k\) from the \(k\)-th layer can be defined as:
\begin{equation}
    A_k=Iterating_k([A_{k-1},T_n])=GA([SA(A_{k-1}),T_n]),
\end{equation}

Generally, the iterating module can be concluded as utilizing attended text features to attend the audio segment embeddings so as to obtain the final output of the audio features. 
Furthermore, because both the stacking and iterating modules are constructed based on the co-attention mechanism, their effects can be identical when each consists of only a single training layer.

\subsection{Model Training with Contrastive Learning}

We utilize a contrastive learning method during the training process to effectively manage the embedding representations obtained in the previous steps. First of all, to conduct the measurement between audio clips and textual captions, both of their output features from the co-attention module, which are \(F_a\) for the audio representations and \(F_t\) for the textual representations, are projected linearly into a joint multimodal space with a dimensio of \(d\). This can be represented as \(E_a\) and \(E_t\) through the following function:
\begin{equation}
    E_a= L_a (F_a ), \quad E_t= L_t (F_t),
\end{equation}
where \(E_{a} \in \mathbb{R}^{N\times d} \) and \(E_{t} \in \mathbb{R}^{1\times d}\), which are the finalized audio and textual representation in this network. $N$ denotes the batch size and \(L_a\), \(L_t\) represents the linear projection functions for audio and textual section, respectively. Then, for a given audio-text pair (\(A_i\), \(T_k\)), in which \(A_i\) represents the \(i\)-th audio samples for an audio embedding \(E_a\) and \(T_k\) represents the \(k\)-th text captions for a textual embedding \(E_t\), we calculate the cosine similarity for the random audio embedding \(A_i\) for a given patch of \(N\) audio-text pairs and for a specific text embedding \(T_k\) by the formulation given below to enable the contrastive learning process:
\begin{equation}
    s(A_i,T_k)=\frac{A_i\cdot(T_k)^T}{\|A_i\|\|T_k\|}.
\end{equation}

For the selection of loss function, the Normalized Temperature-scaled Cross Entropy Loss (NT-Xent) is widely employed in contrastive learning retrieval models training tasks, which is able to effectively handle the relationship of audio-text pairs. We also adjusted it as the baseline loss function in the corresponding part of our work. The corresponding audio-to-text retrieval loss function \(L_i^{A\to T}\) is shown as follows:
\begin{equation}
    L_i^{A\to T}=-log\frac{exp(s(A_i,T_k)/t)}{\sum_{j=1}^N1_{[j\neq i]}exp\big(s(A_i,T_j)/t\big)},
\end{equation}
where \(N\) denotes the batch size of samples, and \(t\) represents the temperature hyper-parameter. This function aims at maximizing the similarity of similar vectors to become closer to the best loss while making this of negative pair samples closer to 0. In such a way, the model is able to highlight those relevant audio-text pairs during the training process so that better training effects can be acquired.

We conduct similar computation to the text embeddings  \(T_i\) for a specific audio embedding \(A_k\) in an input patch of \(N\) audio-text pairs. The formulation of cosine similarity is shown as follows:
\begin{equation}
    s(T_i,A_k)=\frac{T_i\cdot(A_k)^T}{\|T_i\|\|A_k\|}.
\end{equation}

Therefore, the contrastive loss function \(L_{i}^{T\to A}\) for text-to-audio retrieval tasks can be represented as below:

\begin{equation}
    L_{i}^{T\to A}=-log\frac{exp(s(T_{i},A_{k})/t)}{\sum_{j=1}^{N}1_{[j\neq i]}\exp\bigl(s(T_{i},A_{j})/t\bigr)},
\end{equation}
where the batch size of samples is represented by \(N\), and the temperature hyper-parameter is denoted by \(t\) to scale the range of logits. Therefore, the overall contrastive learning loss \(L\) can be represented below by combining and weighing both of the loss functions above with the parameter \(\lambda\):
\begin{equation}
    L=\frac{1}{N}\sum_{i=1}^{N}\bigl(\lambda L_{i}^{A\to T}+(1-\lambda)L_{i}^{T\to A}\bigr).
\end{equation}

\section{Experiments}

\subsection{Datasets}

We conducted experiments on two publicly available datasets: Clotho\cite{clotho} and AudioCaps\cite{audiocaps}. AudioCaps\cite{audiocaps} is a large-scale dataset containing approximately 50K audio recordings and homologous human-written textual descript pairs, which are collected from the AudioSet dataset. The lengths of the recordings are in the range of 0.5 to 10 seconds.
The Clotho\cite{clotho} dataset collects nearly 5K audio samples with five corresponding content description texts for each sample. The signal samples are from Freesound platform after post-processing and last 15 to 30 seconds long. With the consideration of training and updating requirements, we managed to extract 10-second fragments from original files for the Clotho while clipping the 2-second part for AudioCaps. Furthermore, the batch size for AudioCaps is 64 in this project and the one for Clotho is 32. More details for  both datasets can be found in Table \ref{tab:dataset}.

\begin{table}[t]
        \centering
        \caption{Statistics of the Datasets}
        \label{tab:dataset}
        \begin{tabular}{l|llll}
            \toprule
            \multirow{2}{*}{\textbf{Dataset}} & \multirow{2}{*}{\textbf{ Audios}}& \multirow{2}{*}{\textbf{ Text}} & \textbf{Total}  & \textbf{Avg.}   \\ 
            &&& \textbf{duration (h)} & \textbf{durange (s)} \\\midrule %& Batch
            AudioCaps & 49838 & 49838 & 126.7 & 0.5-10 (Short) \\ %& 64 \\
            Clotho & 3840 & 19195 & 24.0 & 15-30 (Long)
            \\\bottomrule % & 32 \\ \hline
        \end{tabular}
    \end{table}

    \begin{table}[t!]
        \centering
        \caption{Performance Comparison on Clotho and AudioCaps}
        \label{tab:comp_perf}
        \resizebox{\linewidth}{!}{\begin{tabular}{c|c|cccc}
        \toprule
        \textbf{Datasets}  & \textbf{Module}            & \textbf{mAP@10}        & \textbf{R@1}           & \textbf{R@5}          & \textbf{R@10}          \\ \midrule
        \multirow{5}{*}{\textbf{Clotho}}    
                  & Baseline  & 24.2  & 13.6  & 35.4  & 47.7 \\
                  & GPTtar \cite{cscwd2024_fuyu}           & 26.5          & 14.3          & 38.4          & 50.2          \\\cmidrule{2-6}
                  & GPTtar+Single             & 29.1          & 16.4          & 40.7          & 51.1          \\
                  & GPTtar+Stacking           & 30.6          & 17.6          & 41.8          & 52.6          \\
                  & GPTtar+\textbf{Iterating} & \textbf{30.9} & \textbf{17.9} & \textbf{42.1} & \textbf{53.0} \\ \midrule
        \multirow{5}{*}{\textbf{AudioCaps}} 
                 & Baseline  & 27.7  & 34.1  & 47.9  & 67.5 \\
                 & GPTtar \cite{cscwd2024_fuyu}           & 32.5          & 36.1          & 63.5          & 76.3 \\\cmidrule{2-6}
                  & GPTtar+Single             & 35.2          & 39.4          & 68.6          & 79.0          \\
                  & GPTtar+Stacking           & 37.1          & 41.3          & 69.7          & 81.3          \\
                  & GPTtar+\textbf{Iterating} & \textbf{37.4} & \textbf{41.5} & \textbf{70.5} & \textbf{82.2} \\ \bottomrule
        \end{tabular}}
    \end{table}

\subsection{Experimental Settings}

In this work, we followed the experiment settings in \cite{cscwd2024_fuyu}, where RoBERTa is adopted as the text encoder and CLAP is adopted as the audio encoder for all experiments. 
The depth of both stacking and iterating network layers is set to 5. For evaluation, we used recall at rank $k$ (R@$k$, $k$=1, 5, and 10) and mean average precision (mAP) to analyze the performance of the models. Similar to \cite{cscwd2024_fuyu}, we  utilized ChatGPT 
to generate five more new sentences for each captions.

\subsection{Results}
The performance of the proposed co-attention methods are compared one baseline method and one state-of-the-art methods: 
\begin{itemize}
    \item \textbf{Baseline} is the baseline method provided in DCASE2023\footnote{https://dcase.community/challenge2023/task-language-based-audio-retrieval}, where we used RoBERTa and CLAP to encode the text and audio. 
    \item \textbf{GPTtar} \cite{cscwd2024_fuyu} is a state-of-the-art method for language-based audio retrieval task, where GPT is used to augment the captions and self attention is used to capture frame-level audio features.
\end{itemize}
The results on different co-attention networks against other methods on the Clotho and AudioCaps datasets is shown in Table \ref{tab:comp_perf}.  From Table \ref{tab:comp_perf}, we can find that: 1) all methods using co-attention networks shows better performance, indicating that the integration of the co-attention module is of better benefit for the network to construct the interaction of different modalities and learn correct features from them;
2) compared with the single network, the stacking and iterating structure for the co-attention units achieve higher accuracy,
showing that with the increase of the network depth, the capabilities of combining and aligning embeddings from audio and textual sections has been improved;
3) the performance of using iterating module is better than that of stacking one, this is mainly because that the learned representation from one modality in the early stage is not accurate enough to incorporate with another modality.

\section{Conclusion}
This paper proposes a novel framework for the language-based audio retrieval task using the co-attention mechanism. Specifically, we combine self-attention and guided-attention unit to form the co-attention module, which aggregates both audio and textual representation in a shared semantic space, thus improving the effects of the training process. Furthermore, stacking and iterating modules are designed by cascading the single co-attention unit to increase the depth of the network and improve the performance.
According to the experiment results, our methods shown better effectiveness in handling language-based audio retrieval task. With increasing the depth of the network, better performance can be obtained. 
We conclude that the application of co-attention mechanism is of help to improve the performance for language-based audio retrieval.

\section*{Acknowledgment}
We would like to acknowledge the support provided by the XJTLU AI University Research Centre, Jiangsu Province Engineering Research Centre of Data Science and Cognitive Computation at XJTLU 
and SIP AI innovation platform (YZCXPT2022103) 
and is also supported by the Research Development Funding (RDF) (contract number RDF-21-02-044) at Xi’an Jiaotong-Liverpool University.

\bibliographystyle{ieeetr}  % Choose a bibliography style
\bibliography{ref}

\end{document}